\documentclass[11pt,showpacs,preprintnumbers,amsmath,amssymb,prd,nofootinbib,superscriptaddress]{revtex4-2}

\usepackage{dcolumn}
\usepackage{bm}
\usepackage{ifpdf}
\usepackage{hyperref}
\usepackage{xcolor,color,graphicx,graphics,physics}
\usepackage[spanish,english]{babel}
\usepackage[latin1]{inputenc}
\usepackage[OT1]{fontenc}
\usepackage{latexsym,amssymb,amsmath,amsfonts, slashed,cancel}
\usepackage{makeidx}
\usepackage{epsfig,subfigure}
\usepackage{natbib}
\usepackage{epstopdf}
\usepackage{mathrsfs}
\usepackage{hyperref}
\hypersetup{colorlinks=true, linkcolor=blue, citecolor=blue, urlcolor=blue}
\usepackage{enumerate}

\usepackage{fixmath}

\everymath{\displaystyle}
\usepackage{graphicx}

\usepackage[T1]{fontenc}
\usepackage{amsmath}
\usepackage{amssymb}
\usepackage{graphicx}
\usepackage{xcolor}

\newcommand{\bea}{\begin{eqnarray}}
\newcommand{\eea}{\end{eqnarray}}

\newcommand{\orcid}[1]{\href{https://orcid.org/#1}{\includegraphics[width=10pt]{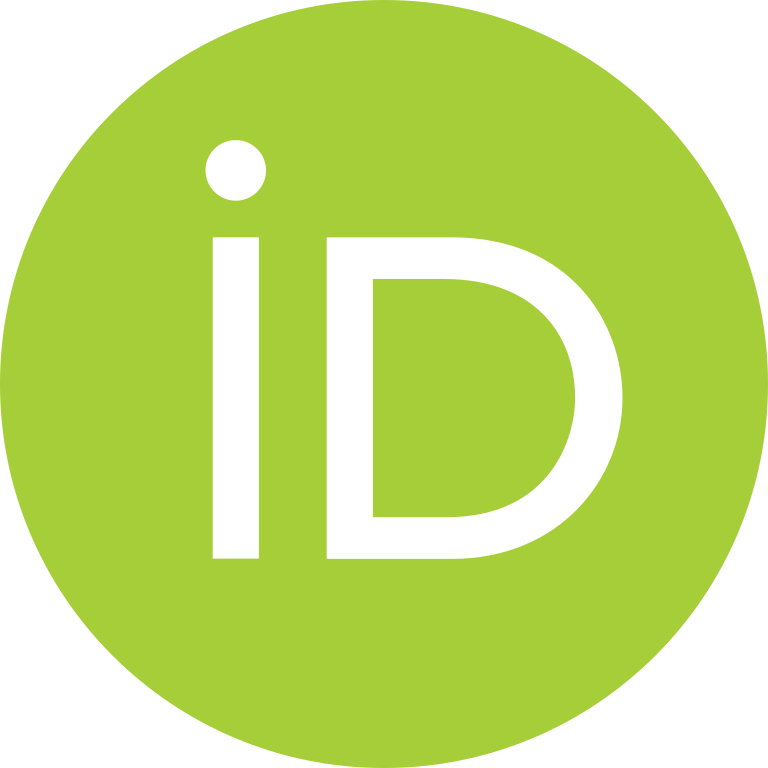}}}

\newcommand{\pr}[1]{\ensuremath{\left[#1\right]}}
\newcommand{\pc}[1]{\ensuremath{\left(#1\right)}}

\newcommand{\R}{R_{\mu\nu}}

\newcommand{\lm}{L_m}

\begin{document}

\title{On causality and its violation in $f(R,\lm,T)$  gravity}

\author{J. S. Gon\c{c}alves \orcid{0000-0001-6704-2748}}
\email{junior@fisica.ufmt.br}
\affiliation{Programa de P\'{o}s-Gradua\c{c}\~{a}o em F\'{\i}sica, Instituto de F\'{\i}sica,\\ 
Universidade Federal de Mato Grosso, Cuiab\'{a}, Brasil}

\author{A. F. Santos \orcid{0000-0002-2505-5273}}
\email{alesandroferreira@fisica.ufmt.br}
\affiliation{Programa de P\'{o}s-Gradua\c{c}\~{a}o em F\'{\i}sica, Instituto de F\'{\i}sica,\\ 
Universidade Federal de Mato Grosso, Cuiab\'{a}, Brasil}

\begin{abstract}

In this paper, $f(R,\lm, T)$ gravity is considered. It is a generalization of the theories $f(R,T)$ and $f(R, \lm)$. This modified theory of gravity exhibits strong geometry-matter coupling. The problem of causality and its violation is verified in this model. Such analysis is carried out using G\"{o}del-type solutions considering different types of matter. It is shown that this model allows both causal and non-causal solutions. These solutions depend directly on the content of matter present in the universe. For the non-causal solution, a critical radius is calculated, beyond which causality is violated. Taking different matter contents, an infinite critical radius emerges that leads to a causal solution. To obtain a causal solution, a natural relationship arises between the parameters of the theory.

\end{abstract}

\maketitle

\section{Introduction}

The discovery of the accelerated expansion of the universe in the late 1990s \cite{Riess, expansao_ace2, Knop, Ama, Wein} led to intense investigation in the search to explain this phenomenon, since a consistent explanation cannot be addressed by general relativity. To explain this late cosmic acceleration, two proposals have been explored:  (i) add new components of matter and energy to the theory of general relativity, namely, dark constituents of the universe; (ii) modify or generalize the Einstein-Hilbert action considering that gravitational action is an arbitrary function that may depend on the geometry and/or of the matter. In this work, the theory under investigation is $f(R, L_m, T)$ gravity \cite{harko}.

The $f(R, L_m, T)$ gravity is a class of modified gravity theories developed to explain the late-time accelerated expansion of the universe, often referred to as the "dark coupling approach". Two additional approaches exist: (i) the dark components model, which extends Einstein's field equations by incorporating two extra terms into the total energy-momentum tensor of the universe, representing dark energy and dark matter separately; and (ii) the dark gravity framework, which offers a purely geometric interpretation of gravitational phenomena. The key idea behind this framework, i.e. dark coupling approach, is to replace the standard Hilbert-Einstein Lagrangian density - which treats the curvature and matter contributions separately and additively - with a more general algebraic formulation that couples these components more intricately. In this model, the Lagrangian is an arbitrary function of the Ricci scalar ($R$), the matter Lagrangian ($L_m$), and the trace of the energy-momentum tensor ($T$). This approach naturally introduces a non-minimal interaction between matter and spacetime geometry, allowing for various forms of coupling, such as between the Ricci scalar and the trace of the energy-momentum tensor.

In this context, $f(R, L_m, T)$ gravity unifies two distinct modified gravity models: $f(R, T)$ and $f(R, L_m)$. Moreover, it introduces a non-minimal coupling of matter to geometry through both the matter Lagrangian and the trace of the energy-momentum tensor. A significant feature of this theory is that the resulting modifications apply non-trivially to all types of matter. As limiting cases this theory recovers the $f(R)$, $f(R, L_m)$ and $f(R, T)$ gravity theories. It is important to emphasize that all alternative theories proposed as gravitational theory must be intensively tested. A classical test consists of checking whether all exact solutions of general relativity are contained in this new model. Here causality and its violation are investigated.

To address the challenge of causality within the domain of $f(R,\lm, T)$ gravity, a comprehensive investigation is performed using the G\"{o}del-type solution \cite{tipo_godel1}. This metric is an extension of Kurt G\"{o}del solution, proposed in 1949 \cite{godel}. It represents an exact solution that incorporates rotating matter within the domain of general relativity. The main characteristic of this metric is its ability to accommodate Closed Timelike Curves (CTCs). CTCs are curves that lead to causality violation. A traveler on this curve can return to their past, at least theoretically. However, it is important to note that this is a global point of view since locally the general relativity satisfies the same conditions as the special theory of relativity, that is, the causality is preserved. On the other hand, CTCs and causality violation are not exclusive to G\"{o}del-type solutions, this is a phenomenon whose presence extends beyond the limits of the G\"{o}del solution. These CTCs manifest in a variety of cosmological models, such as the Kerr black hole \cite{ctc2}, Van-Stockum model \cite{Van}, cosmic strings \cite{ctc1}, among others. To further explore causality violation, the G\"{o}del-type metric allows the calculation of a critical radius $r_c$, a parameter that demarcates regions of causal and non-causal behaviors.

As the study of causality is a classical test for alternative models of gravitation, the G\"{o}del and type-G\"{o}del solutions have been studied in several modified gravity theories. Such an investigation took place in $f(R)$ theory \cite{fr_and_godel}, in $k$-essence theory \cite{kessence_and_godel}, in Chern-Simons gravity \cite{chersimon_and_godel1, chersimon_and_godel2}, in $f(T)$ gravity \cite{ft_and_godel}, in $f(R,T)$ gravity \cite{frt_and_godel}, in bumblebee gravity \cite{bumblebeee_and_godel}, in Horava-Lifshitz gravity \cite{horava_and_godel}, in Brans-Dicke theory \cite{brans_and_godel}, in $f(R,Q)$ gravity \cite{frq_and_godel}, in $f(R,T)$ Palatini gravity \cite{frt_palatini}, in $f(R,\phi,X)$ \cite{fr_phi_X}, among others. In this paper, the main objective is to investigate whether $f(R,\lm, T)$ gravity permits a G\"{o}del-type solution, which is an exact solution in general relativity. Specifically, we aim to explore if this new gravitational model can accommodate such a solution. Different types of matter content are considered, and the analysis is divided into four cases: (i) a perfect fluid; (ii) an electromagnetic field; (iii) a combination of a perfect fluid and an electromagnetic field; and (iv) a scalar field. Within this framework, both causal and non-causal regions are identified. In each case, a direct comparison with the corresponding general relativity solution is made to evaluate how $f(R,\lm, T)$ gravity modifies or preserves the G\"{o}del-type behavior.

The present paper is organized as follows. In Section II, $f(R,\lm, T)$ gravity is introduced and the field equations are derived. In Section III, the G\"{o}del-type metric is discussed. In order to solve the field equations, different types of matter content are considered. Taking a perfect fluid as matter content the standard G\"{o}del solution is obtained. Then causality violation is allowed. A critical radius is calculated and the influence of $f(R,\lm, T)$ gravity on this quantity is displayed.  Considering an electromagnetic field, a combination of a perfect fluid plus an electromagnetic field and a scalar field as matter content, causal solutions are found. Furthermore,  to obtain a causal solution a new restriction is required. In Section IV, we investigate a specific case of $f(R,\lm)$ gravity, referred to as exponential $f(R,\lm)$ gravity, considering three types of matter content. In Section V, some remarks and conclusions are presented.

\section{$f(R,\lm, T)$ Gravity}

In this section, the $f(R,\lm, T)$ theory is presented and its field equations are obtained. The action that describes this gravitational theory is defined as
\begin{equation}
    S= \frac{1}{16\pi} \int f(R,\lm, T) \sqrt{-g} \  d^4x + \int \lm \sqrt{-g} \ d^4x,  
\end{equation}
where $f(R,\lm, T)$  is an arbitrary function of the Ricci scalar $R$, of the trace of the energy-momentum tensor $T$, with $T = g^{\mu\nu}T_{\mu\nu}$, and of the Lagrangian density of ordinary matter, $\lm$, respectively. It is important to emphasize that the energy-momentum tensor is defined according to
the relation:
\begin{equation}\label{tensor energia}
        T_{\mu\nu} = \frac{-2}{\sqrt{-g}} \frac{\partial \pc{\sqrt{-g}\lm}}{\partial g^{\mu\nu}} = -2 \frac{\partial \lm}{\partial g^{\mu\nu}} + g_{\mu\nu} \lm.
\end{equation}

In order to obtain the field equations, the variation of the action with respect to the metric tensor is taken. Then
\begin{equation}
    \delta S = \frac{1}{16\pi} \int d^4x \pr{\delta \sqrt{-g} f + \sqrt{-g} \pc{f_R \delta R + f_T\delta T + f_L \delta L_m }} + \int d^4x \delta \pc{\sqrt{-g} \lm}\label{eq3}
\end{equation}
with $f_R \equiv \frac{\partial f}{\partial R}$, $f_T \equiv \frac{\partial f}{\partial T}$ and $f_L \equiv \frac{\partial f}{\partial \lm}$. The variations $\delta R$ and $\delta T$ are known \cite{reviewbook}. They are given as
\begin{equation}
    f_R \delta R + f_T \delta T = \pr{\R f_R + \pc{g_{\mu\nu} \Box - \nabla_ \mu \nabla_ \nu}f_R + \pc{T_{\mu\nu} + \Theta_{\mu\nu}}f_T} \delta g^{\mu\nu},
\end{equation}
where the tensor $\Theta_{\mu\nu}$ is defined based on the variation
\begin{align}
\delta T &= \delta \pc{g^{\alpha \beta} T_{\alpha\beta}},\\
&= \pc{\frac{\delta g^{\alpha\beta}}{\delta g^{\mu\nu}}} T_{\alpha\beta} + \frac{\delta T_{\alpha\beta}}{\delta g^{\mu\nu}} g^{\alpha\beta},\\
&=\pc{T_{\mu\nu} + \Theta_{\mu\nu}}, 
\end{align}
which leads to
\begin{equation}
    \Theta_{\mu\nu} \equiv \frac{\delta T_{\alpha\beta}}{\delta g^{\mu\nu}} g^{\alpha\beta} = -2T_{\mu\nu} + g_{\mu\nu} \lm  - \tau_{\mu\nu}.
\end{equation}
Here a new tensor $\tau_{\mu\nu}$ is defined as 
\begin{equation}
    \tau_{\mu\nu} = 2 g^{\alpha\beta} \frac{\partial^2 \lm}{\partial g^{\mu\nu}\partial g^{\alpha\beta}}. 
\end{equation}

In addition, the term $f_L \delta L_m$ in Eq. (\ref{eq3}) becomes
\begin{equation}
    f_L \delta L_m = \frac{1}{2} f_L (g_{\mu\nu} L_m - T_{\mu\nu}). 
\end{equation}
Taking these results, the complete set of the field equations is given by
\begin{equation}\label{eq_campo}
     f_R R_{\mu\nu} -\frac{1}{2} \pr{f-(f_L + 2f_T)\lm} g_{\mu\nu} + (g_{\mu\nu} \Box - \nabla_ \mu \nabla_ \nu)f_R = \pr{\kappa + \frac{1}{2} (f_L + 2 f_T)} T_{\mu\nu} + f_T \tau_{\mu\nu},
\end{equation}
where $\kappa=16\pi$ is a constant.

The main objective of this work is to investigate whether the causality violation described by G\"{o}del-type universes persists in this gravitational theory as allowed in general relativity.

\section{G\"{o}del-type Metric}

In this section, the proposal is to present the G\"{o}del-type solution and analyze whether this universe is a solution of $f(R,\lm, T)$ gravity. The study will be developed for different contents of matter. The G\"{o}del-type solution represents a generalization of the G\"{o}del metric. The G\"{o}del universe is an exact solution to Einstein field equations for a homogeneous rotating universe. This solution introduces the possibility of Closed Time-like Curves (CTCs), which allow for causality violations. Consequently, this space-time allows theoretical time travel. 
 
In order to obtain more details about causality violation, the G\"{o}del-type metric is considered \cite{tipo_godel1}. The line element that describes this universe is given by 
\begin{equation}\label{type_godel}
    ds^2 = -dt^2 - 2H(r) dtd\phi + dr^2 + \left(D^2(r)-H^2(r)\right) d\phi^2 + dz^2,
\end{equation}
where $H(r)$ and $D(r)$ are functions that satisfies the relations
\bea
\frac{H'(r)}{D(r)}=2\omega \quad \mathrm{and}\quad \frac{D''(r)}{D(r)}=m^2.
\eea
The prime ($'$) denotes the derivative with respect to $r$. 

It is worth noting that the parameters $\omega$ and $m^2$ are free parameters that characterize all the properties of the metric. Theses parameters define three distinct classes: hyperbolic ($m^2>0, \omega \neq 0$), trigonometric ($m^2<0, \omega \neq 0$), and linear ($m^2=0, \omega \neq 0$) \cite{tipo_godel1}.  Here only the hyperbolic class is considered, since this class contains the standard G\"{o}del universe. For this class, the functions $H(r)$ and $D(r)$ are given as
\begin{align}
    H(r) &= \frac{4\omega}{m^2} senh^2 \left(\frac{mr}{2} \right),\label{condition1} \\
     D(r) &= \frac{1}{m} senh(mr)\label{condition2}.
\end{align}

The critical radius, which defines the existence of Closed Time-like Curves (CTCs), is an important quantity that can be derived from the G\"{o}del-type solution. It is defined as
\begin{equation}
r_c = \frac{2}{m} \sinh^{-1} \left(\frac{4\omega^2}{m^2}-1\right)^{-1}.\label{cr}
\end{equation}
This expression leads to two interesting relationships between the parameters $\omega^2$ and $m^2$. Firstly, when $m^2=2\omega^2$, the G\"{o}del solution exhibits a finite critical radius given by
\begin{equation}
r_c = \frac{2}{m} \sinh^{-1} \left(1\right).
\end{equation}
Consequently, non-causal regions are allowed. Secondly, if $m^2=4\omega^2$, the critical radius becomes infinite. In other words, this condition results in a causal solution.

For simplicity, a new basis, known as a local Lorentz co-frame, is chosen, transforming the metric into the form \cite{tipo_godel1},
\begin{equation}
ds^2 = \eta_{AB} \theta^A \theta^B =(\theta^{(0)})^2 - (\theta^{(1)})^2 -(\theta^{(2)})^2 - (\theta^{(3)})^2, \label{frame}
\end{equation}
where $\eta_{AB}$ represents the Minkowski metric, and $ \theta^A = e^A\ _{\mu}\ dx^\mu$ with
\begin{align}
\theta^{(0)} &= dt + H(r)d\phi,\nonumber \\ \theta^{(1)} &= dr, \nonumber\ \\ \theta^{(2)} &= D(r)d\phi,\nonumber \\ \theta^{(3)} &= dz. \label{coframe}
\end{align}
Here, $e^A\ {\mu}$ corresponds to the tetrad that satisfies the relation $ e^A\,_\mu e^\mu\,_B=\delta^A_B$. The capital Latin letters are used to label Lorentz indices ranging from 0 to 3.

In the local Lorentz co-frame (\ref{coframe}), the non-zero Ricci tensor components are  $R_{(0)(0)} = 2\omega^2, \ R_{(1)(1)} = \ R_{(2)(2)}= 2\omega^2 - m^2.$ The Ricci scalar is $R = 2(m^2 - w^2)$. 

Now, the objective is to use these quantities and solve the Einstein field equations (\ref{eq_campo}) choosing different matter contents.

\subsection{Perfect Fluid}

In this subsection, the G\"{o}del-type metric is analyzed in $f(R,\lm, T)$ gravity considering the perfect fluid as the matter content. The energy-momentum tensor that describes the perfect fluid is given as
\begin{eqnarray}
    T_{AB}=(\rho+p)u_A u_B-p\eta_{AB},\label{EMT}
    \end{eqnarray}
with $\rho$ being the energy density, $p$ the pressure and $u_A = (1,0,0,0)$ the four-velocity of the fluid. The non-zero components of the energy-momentum tensor are $T_{00} = \rho ,  T_{11} = T_{22} = T_{33} = p $. The Lagrangian that describes the perfect fluid is defined as $L_M = -p$.

The field equation Eq. (\ref{eq_campo}) in tangent space becomes
\begin{equation}\label{eq_campo_tangente}
     f_R R_{AB} -\frac{1}{2} \pr{f-(f_L + 2f_T)\lm} \eta_{AB} + (\eta_{AB} \Box - \nabla_ A \nabla_B)f_R = \pr{\kappa + \frac{1}{2} (f_L + 2 f_T)} T_{AB} + f_T \tau_{AB}.
\end{equation}
Taking into account the perfect fluid, the tensor $\tau_{AB}$ becomes zero. In addition, the derivatives of the function $f_R$ are zero, since the Ricci scalar is constant. Therefore, the field equations are given as
\begin{align}
     2 f_R \omega^{2} - \frac{1}{2} f &= \rho \kappa + (\rho + p)(\frac{1}{2} f_L +  f_T ),\label{m}\\
     (2\omega^2 - m^2)f_R + \frac{1}{2} f &= \kappa p \label{eq1_campo},\\
       \frac{1}{2} f &=  \kappa p\label{eq2_campo}.
\end{align}
Eqs. (\ref{eq1_campo}) and (\ref{eq2_campo}) lead to $m^2 = 2\omega^2$. This condition defines the G\"{o}del solution. Then the causality violation is allowed in this gravitational theory. 

The remaining equation Eq. (\ref{m}) provides
\begin{equation}
    m = \pr{\frac{(\rho + p)(\frac{1}{2}f_L + f_T) +\frac{1}{2}f + \kappa \rho }{f_R}}^{1/2}.
\end{equation}
Then the critical radius in $f(R,T, \lm)$ gravity is 
\begin{eqnarray}
    r_c=2\sinh^{-1}(1)\pr{\frac{f_R}{(\rho + p)(\frac{1}{2}f_L + f_T) +\frac{1}{2}f + \kappa \rho } }^{1/2}.
\end{eqnarray}

This result shows that the critical radius, beyond which the causality violation explicitly occurs, depends on the form of the function $f(R, \lm, T)$ and its derivatives,
$f_R, f_L$ and $f_T$. This result is a generalization of the study developed for $f(R)$ and $f(R,T)$ gravity.

In short, assuming the perfect fluid as the unique matter of the universe, the causality violation arises naturally. Now the question is: do other types of matter lead to the same consequence? Let us carry out this investigation in the next subsections.

\subsection{Electromagnetic Field}

In an attempt to find a causal solution for the G\"{o}del-type metric in $f(R, \lm, T)$ gravity the electromagnetic field is considered as the content of matter. For simplicity, is assumed that the electromagnetic field is aligned on the $z$-axis and depends on $z$  \cite{tipo_godel1}. For this choice, the non-vanishing components of the electromagnetic tensor in the flat space-time are
\begin{equation}
F_{(0)(3)}=-F_{(3)(0)}=E(z)\quad\mathrm{e}\quad F_{(1)(2)}=-F_{(2)(1)}=B(z),\label{27}
\end{equation}
where $E(z)$ and $B(z)$ are solutions of Maxwell equations given by
\begin{align}
    E(z)&=E_0\cos\left[2\omega(z-z_0)\right],\label{28}\\
    B(z)&=E_0\sin\left[2\omega(z-z_0)\right],\label{29}
\end{align}
with $E_0$ being the amplitude of the electric and magnetic fields and $z_0$ is an arbitrary constant. Thus, the non-zero components of the energy-momentum tensor associated with the electromagnetic field are given by 
\begin{equation}\label{energy_sol}
    T_{(0)(0)}^{EM} = T_{(1)(1)}^{EM} = T_{(2)(2)}^{EM} = \frac{E_0^2}{2}, \quad\quad T_{(3)(3)}^{EM} = - \frac{E_0^2}{2}.
\end{equation}

For this type of choice, the Lagrangian that describes the electromagnetic field is 
\bea
L_M = -\frac{1}{4}F_{AB}F^{AB} = \frac{1}{2} (B^2 - E^2).
\eea
As a consequence $\tau_{AB} = F_{CA} F_B \ ^C$ whose non-zero components are
\begin{equation}
    \tau_{33} = -\tau_{00} = E^2 \quad\mathrm{and}\quad \tau_{11} = \tau_{22} = -B^2. 
\end{equation}
Then, the field equations (\ref{eq_campo_tangente}) become
\begin{align}
    4\omega^2f_R - f &= -\frac{1}{2}(B^2-E^2) ( f_L + 2f_T) + E_0^{2} \pr{\kappa + \frac{1}{2} (f_L +2f_T)} -2E^{2} f_T, \\
    2(2\omega^2 - m^2)f_R + f &= \frac{1}{2}(B^2-E^2) ( f_L +2f_T) + E_0^{2} \pr{\kappa + \frac{1}{2} (f_L + 2f_T)} - 2B^2 f_T,  \label{eq2eletro} \\
   - f &= - \frac{1}{2}(B^2-E^2) (f_L + 2f_T) + E_0^{2} \pr{\kappa + \frac{1}{2}(f_L + 2f_T) }- 2E^{2} f_T . \label{eq3eletro}
\end{align}

Eqs. (\ref{eq2eletro}) and (\ref{eq3eletro}) lead to
\begin{equation}
    (m^2 - 2\omega^2) f_R =  -E_0^{2} \pr{\kappa + \frac{1}{2} (f_L +2f_T)} + f_T(E^2 + B^2).
\end{equation}
Considering that $f_R > 0$ and 
\begin{equation}
    f_T(E^2 + B^2) > E_0^{2} \pr{\kappa + \frac{1}{2} (f_L +2f_T)},
\end{equation}
a causal G\"{o}del-type solution, i.e., $m^2 = 4 \omega^2$ is allowed. This implies that the critical radius, defined in Eq. (\ref{cr}), becomes infinity. Therefore, it is possible a causal G\"{o}del-type solution in $f(R, \lm, T)$ gravity when the matter content is an electromagnetic field.

\subsection{Perfect Fluid plus Electromagnetic Field}

Here let us investigate the consistency of the G\"{o}del-type universe in $f(R, \lm, T)$ gravity considering as matter content a combination of perfect fluid and eletromagnetic field. The energy-momentum tensor associated with this combination of matter is written as follows
\begin{equation}
    T_{AB}=T_{AB}^{PF}+T_{AB}^{EM},
\end{equation}
where $T_{AB}^{PF}$ and $T_{AB}^{EM}$ are energy-momentum tensors of the perfect fluid and electromagnetic field, respectively. Then, the non-zero components of the total energy-momentum tensor are
\begin{align}
    T_{00} &= \frac{E_{0}^2}{2} + \rho, \\ T_{11} &= T_{22} =\frac{E_{0}^2}{2} - p, \\  T_{33} &= - \frac{E_{0}^2}{2}  -p.
\end{align}

Using these components the field equations (\ref{eq_campo_tangente}) are written as
\begin{align}
    4\omega^2 f_R - f &= -\pr{\frac{1}{2} (B^2 - E^2) - p} (f_L + 2f_T) + 2\pc{\rho + \frac{E_0^{2}}{2}} \pr{\kappa + \frac{1}{2}(f_L + 2f_T)} - 2E^2 f_T, \\
    2(2\omega^2 -m^2)f_R + f &=  \pr{\frac{1}{2}(B^2 - E^2) - p}(f_L + 2f_T) + 2 \pc{\frac{E_0^{2}}{2} - p}\pr{\kappa + \frac{1}{2}(f_L + 2f_T)}  - 2B^2 f_T, \label{2plus} \\
    -f &= - \pr{\frac{1}{2} (B^2 - E^2) - p}(f_L + 2f_T) + 2 \pc{\frac{E_0^{2}}{2} + p} \pr{\kappa + \frac{1}{2}(f_L + 2f_T)} - 2E^2 f_T. \label{3plus}
\end{align}

Adding the equations (\ref{2plus}) and (\ref{3plus}), we obtain
\begin{equation}
    (m^2 - 2\omega^2 )f_R = - E_0^{2}  \pr{\kappa + \frac{1}{2}(f_L + 2f_T)} + f_T(E^2 + B^2).
\end{equation}

This equation leads to the following interpretation. Assuming that $f_R > 0$, and
\begin{equation} \label{condition}
    - E_0^{2}  \pr{\kappa + \frac{1}{2}(f_L + 2f_T)} + f_T(E^2 + B^2) > 0,
\end{equation}
the causal solution $m^2 = 4\omega^2$ is allowed in this gravitational theory. As a consequence $r_c \rightarrow \infty$. In addition, substituting the values of the electric and magnetic fields, given in Eqs. (\ref{28}) and (\ref{29}),  in Eq. (\ref{condition}), a new restriction emerges, i.e.
\begin{align}
    f_L < -2\kappa. \label{new_result}
\end{align}

Therefore, for this choice of matter content, perfect fluid plus electromagnetic field, a causal G\"{o}del-type universe is solution of $f(R, \lm, T)$ gravity. In this particular case, a restriction is imposed on the derivative of the function $f(R,\lm, T)$ with respect to the Lagrangian of matter. It is important to note that the same condition for $f_L$ also appears for the case where the matter content is just the electromagnetic field.


\subsection{Scalar field}

In this subsection, the main objective is to find a causal solution within the $f(R, L_m, T)$ gravity theory, considering a different type of matter content. In this case, a massless scalar field is chosen. The Lagrangian describing the scalar field $\Phi$ is given as
\begin{equation}
    \mathcal{L} = \eta^{AB} \nabla_A \Phi \nabla_B \Phi. \label{48}
\end{equation}
The corresponding energy-momentum tensor reads
\begin{equation}
T_{AB}=\nabla_A \Phi \nabla_B \Phi - \frac{1}{2} \eta_{AB}\eta^{CD}\nabla_C \Phi \nabla_D \Phi.
\end{equation}
Assuming that $\Phi = \epsilon z + \epsilon $, with $\epsilon = const$, the Lagrangian (\ref{48}) becomes
\begin{equation}
   \mathcal{L} =- \epsilon^2,
\end{equation}
and the non-zero components of the energy-momentum tensor are
\begin{equation}
    T_{00} = -T_{11} = -T_{22}= T_{33} = \frac{\epsilon^2}{2}.
\end{equation}
The trace of $T_{AB}$ is given by
\begin{equation}
    T=\epsilon^2.
\end{equation}
As discussed in \cite{harko}, the tensor $\tau_{AB}$ is zero for the scalar field.

With these ingredients, the field equations  (\ref{eq_campo_tangente}) generate the set of equations
\bea
2\omega^2 f_R-\frac{1}{2}f&=&\frac{\kappa\epsilon^2}{2}+\frac{3\epsilon^2}{4}(f_L+2f_T),\label{53}\\
(2\omega^2-m^2) f_R+\frac{1}{2}f&=&-\frac{\kappa\epsilon^2}{2}-\frac{3\epsilon^2}{4}(f_L+2f_T),\label{54}\\
\frac{1}{2}f&=&\frac{\kappa\epsilon^2}{2}-\frac{\epsilon^2}{4}(f_L+2f_T).\label{55}
\eea
Working with the Eqs. (\ref{53}) and (\ref{54}) we obtain
\bea
m^2=4\omega^2.
\eea
This solution implies that the critical radius becomes infinite. Then a causal solution is allowed.
In addition, Eqs. (\ref{54}) and (\ref{55}) lead to
\bea
(m^2-2\omega^2)=\kappa\epsilon^2+\frac{\epsilon^2}{2}(f_L+2f_T).
\eea
A causal class of solution appears from this equation assuming that $f_L>0$ and $f_T>0$.
 Therefore, assuming the scalar field as the content of matter, $f(R, \lm, T)$ gravity allows for a causal G\"{o}del-type solution. In other words, for this content of matter, no causality violation is permitted. It is important to note that if a combination of a perfect fluid plus a scalar field is taken as the matter content, similar results will be obtained. Furthermore, the results presented in this manuscript are a generalization of the study developed in references \cite{fr_and_godel, frt_and_godel} for $f(R)$ gravity and $f(R, T)$ gravity, respectively.

It should be noted that the study developed in this work does not investigate the implications for the conservation or non-conservation of the energy-momentum tensor. However, as discussed in \cite{harko}, in $f(R, \lm, T)$ gravity, the matter energy-momentum tensor $T_{\mu\nu}$ is generally not conserved. Since this non-conservation of the energy-momentum tensor can have physical implications, it is important to consider.

In the next section, a particular case is analyzed.

\section{Particular case: exponential $f(R, \lm)$ gravity}

Here a particular case of $f(R, \lm)$ gravity is considered. In this case the action of the gravitational field exhibits an exponential dependence on the standard Hilbert-Einstein Lagrangian density \cite{TH}. The action for this modified theory of gravity takes the following form
\begin{equation}
    S= \int f(R,\lm) \sqrt{-g} \  d^4x , \label{58} 
\end{equation}
where the function $f(R,\lm)$ is given as
\bea
f(R,\lm)=\Lambda \exp\left(\frac{1}{2\Lambda}R+\frac{1}{\Lambda}\lm\right),\label{59}
\eea
with $\Lambda$ being an arbitrary positive constant. The standard general relativity action is recovered in the limit $(1/2\Lambda)R+(1/\Lambda)\lm \ll 0$.

Using the function (\ref{59}) and taking the variation of the action (\ref{58}) with respect to the metric, the gravitational field equations in the local Lorentz co-frame are as follows
\bea
R_{AB}&=&\left(\Lambda-\lm\right)\eta_{AB}+T_{AB}-\frac{1}{\Lambda}\Biggl[\left(\frac{1}{2}\nabla_C\nabla^C R+\nabla_C\nabla^C\lm\right)\eta_{AB}\nonumber\\
&&-\left(\frac{1}{2}\nabla_A\nabla_B R+\nabla_A\nabla_B\lm\right)\Biggl]-\frac{1}{\Lambda^2}\Biggl[\left(\frac{1}{2}\nabla^C R+\nabla^C\lm\right)\left(\frac{1}{2}\nabla_C R+\nabla_C\lm\right)\eta_{AB}\nonumber\\
&&-\left(\frac{1}{2}\nabla_A R+\nabla_A\lm\right)\left(\frac{1}{2}\nabla_B R+\nabla_B\lm\right)\Biggl].
\eea

For the G\"{o}del-type metric, the Ricci scalar has a constant value given by $R=2(m^2-\omega^2)$. As a consequence, we obtain that $\nabla_C\nabla^C R=0$, $\nabla_A\nabla_B R=0$ and $\nabla^C R=0$. Then the field equations become
\bea
R_{AB}&=&\left(\Lambda-\lm\right)\eta_{AB}+T_{AB}-\frac{1}{\Lambda}\left[\nabla_C\nabla^C\lm\eta_{AB}-\nabla_A\nabla_B\lm\right]\nonumber\\
&&-\frac{1}{\Lambda^2}\left[\left(\nabla^C\lm\right)\left(\nabla_C\lm\right)\eta_{AB}-\left(\nabla_A\lm\right)\left(\nabla_B\lm\right)\right].\label{61}
\eea

Since the modified Einstein equations depend only on the matter Lagrangian, let us analyze them by considering different matter contents. 

Assuming that the content of matter is a perfect fluid with Lagrangian $L_m=-p$, or a scalar field with Lagrangian $L_m=\eta^{AB}\partial_A\Phi\partial_B\Phi$, we find that $\nabla_C\nabla^C\lm=0$, $\nabla_A\nabla_B\lm=0$ and $\nabla^C\lm=0$. As chosen before $\Phi = \epsilon z + \epsilon $, with $\epsilon = const$. Therefore, for these matter contents the exponential $f(R, \lm)$ gravity recovers all standard results for the G\"{o}del-type universe.

Now let us choose the electromagnetic field whose Lagrangian is
\bea
L_M = -\frac{1}{4}F_{AB}F^{AB} = \frac{1}{2} (B^2 - E^2),\label{62}
\eea
where the conditions given in Eqs. (\ref{27}), (\ref{28}) and (\ref{29}) have been used. Using Eq. (\ref{energy_sol}) and the Lagrangian (\ref{62}),  the Eq. (\ref{61}) leads to the set of equations
\bea
2\omega^2&=&\Lambda-\frac{1}{2}(B^2-E^2)+\frac{E_0^2}{2}-\frac{8\omega^2(B^2-E^2)}{\Lambda}-\frac{16\omega^2B^2E^2}{\Lambda^2},\label{63}\\
2\omega^2-m^2&=&-\Lambda+\frac{1}{2}(B^2-E^2)+\frac{E_0^2}{2}+\frac{8\omega^2(B^2-E^2)}{\Lambda}+\frac{16\omega^2B^2E^2}{\Lambda^2},\label{64}\\
0&=&-\Lambda+\frac{1}{2}(B^2-E^2)-\frac{E_0^2}{2}+\frac{32\omega^2B^2E^2}{\Lambda^2}.\label{65}
\eea
Eqs. (\ref{64}) and (\ref{65}) lead to
\bea
2\omega^2-m^2=E_0^2+\frac{8\omega^2(B^2-E^2)}{\Lambda}-\frac{16\omega^2B^2E^2}{\Lambda^2}
\eea
This equation implies that the condition $m^2\geq 4\omega^2$ can be satisfied whether
\bea
E_0^2\Lambda^2+8\omega^2(B^2-E^2)\Lambda-16\omega^2B^2E^2<0.
\eea
Therefore, this gravitational theory exhibits a causal G\"{o}del-type behavior, whether the cosmological constant $\Lambda$ obeys the limit.
\bea
\Lambda_1<\Lambda<\Lambda_2
\eea
with 
\bea
\Lambda_1&=&\frac{-b-\sqrt{b^2-4ac}}{2a},\\
\Lambda_2&=&\frac{-b+\sqrt{b^2-4ac}}{2a},
\eea
where the constants are defined as $a=E_0^2$, $b=8\omega^2(B^2-E^2)$ and $c=16\omega^2B^2E^2$. 

This result shows that exponential $f(R, \lm)$ gravity, with the electromagnetic field as the matter content, leads to the non-existence of G\"{o}del's circles and, consequently, avoids the breakdown of G\"{o}del-type causality.

\section{Discussions and Final Remarks}

The present study focuses on $f(R, \lm, T)$ gravity. It is a gravitational theory that generalizes the coupling between geometry and matter. Analyzing the field equations, it is evident that the G\"{o}del-type metric is a solution within this theory. The investigation is carried out for different types of matter. Considering a perfect fluid as matter content, a non-causal solution is identified. This leads to the emergence of a critical radius that delineates a finite non-causal region. This critical radius depends on the parameters of the theory. In the search for a causal solution, an electromagnetic field is incorporated. As a consequence, a causal solution arises, which means that the critical radius tends to the infinity. Similar behavior occurs for a combination of a perfect fluid plus an electromagnetic field as matter content. A causal solution is also found in the case where the matter content is a scalar field. Furthermore, it is important to note that the theories $f(R)$, $f(R, T)$ and $f(R, L_m)$ and their causal or non-causal behavior are recovered in an appropriate limit. In addition, the results obtained here show that to preserve causality in $f(R, L_m, T)$  gravity a specific condition under the parameter $f_L$ is necessary.  Additionally, the results for the modified theory of gravity can be directly compared to those of general relativity, as the G\"{o}del-type solution is also an exact solution within general relativity. While both theories permit causal and non-causal regions, the conditions leading to these regions differ. In $f(R, L_m, T)$ gravity, certain restrictions arise that depend on the parameters of the theory. The main finding - that $f(R, L_m, T)$ gravity allows the same exact solution as general relativity - serves as an important test for this theory, which was constructed to explain gravitational phenomena. Another important aspect to highlight concerns the critical radius. Our results indicate that, when considering different types of matter, the critical radius is finite in the case of a perfect fluid, while it becomes infinite for other forms of matter. Consequently, causal and non-causal, or physical and non-physical, regions are identified, which depend on the choice of matter. Although the critical radius is not a physical quantity in itself, it serves to delineate the physical and non-physical regions within this cosmological framework. Thus, the critical radius demonstrates behavior similar to that observed in general relativity. Another important point to note is that all results are derived for a general form of the function $f(R, L_m, T)$, making our findings applicable to all examples of this function.

A similar study has been conducted for a specific case of $f(R, \lm)$ gravity, known as exponential $f(R, \lm)$ gravity. The analysis considered three different types of matter content: perfect fluid, scalar field, and electromagnetic field. In the first two cases, the same results as general relativity are obtained. However, for the electromagnetic field, an interesting condition emerges: to prevent the violation of causality, the cosmological constant must satisfy a specific limit.

\section*{Acknowledgments}

This work by A. F. S. is partially supported by National Council for Scientific and Technological Develo\-pment - CNPq project No. 312406/2023-1. J. S. Gon\c{c}alves thanks CAPES for financial support. 

\section*{Data Availability Statement}

No Data associated in the manuscript.


\global\long\def\link#1#2{\href{http://eudml.org/#1}{#2}}
 \global\long\def\doi#1#2{\href{http://dx.doi.org/#1}{#2}}
 \global\long\def\arXiv#1#2{\href{http://arxiv.org/abs/#1}{arXiv:#1 [#2]}}
 \global\long\def\arXivOld#1{\href{http://arxiv.org/abs/#1}{arXiv:#1}}


\end{document}